\documentclass[
  journal=largetwo,
  manuscript=article-type,
  year=2024,
  volume=37,
]{cup-journal}

\usepackage{amsmath}
\usepackage{amsfonts}
\usepackage{amssymb}
\usepackage[nopatch]{microtype}
\usepackage{booktabs}
\usepackage[table]{xcolor}
\usepackage{multirow}
\usepackage{aas_macros}

\defcitealias{NFW1995}{NFW 1995}
\defcitealias{NFW1996}{NFW 1996}
\defcitealias{NFW1997}{NFW 1997}

\title{Predicting the Scaling Relations between the Dark Matter Halo Mass and Observables from Generalised Profiles I: Kinematic Tracers}

\author{A. Sullivan}
\affiliation{International Centre for Radio Astronomy Research, The University of Western Australia, 35 Stirling Highway, Crawley, Western Australia, 6009, Australia}
\alsoaffiliation{ARC Centre of Excellence for All Sky Astrophysics in 3 Dimensions (ASTRO 3D)}

\author{C. Power}
\affiliation{International Centre for Radio Astronomy Research, The University of Western Australia, 35 Stirling Highway, Crawley, Western Australia, 6009, Australia}
\alsoaffiliation{ARC Centre of Excellence for All Sky Astrophysics in 3 Dimensions (ASTRO 3D)}

\author{C. Bottrell}
\affiliation{International Centre for Radio Astronomy Research, The University of Western Australia, 35 Stirling Highway, Crawley, Western Australia, 6009, Australia}

\email[A. Sullivan]{andrew.sullivan@icrar.org}


\begin{document}

\begin{abstract}
	We investigate the relationship between a dark matter halo's mass profile and measures of the velocity dispersion of kinematic tracers within its gravitational potential. By predicting the scaling relation of the halo mass with the aperture velocity dispersion, $M_\mathrm{vir} - \sigma_\mathrm{ap}$, we present the expected form and dependence of this halo mass tracer on physical parameters within our analytic halo model: parameterised by the halo's negative inner logarithmic density slope, $\alpha$, its concentration parameter, $c$, and its velocity anisotropy parameter, $\beta$. For these idealised halos, we obtain a general solution to the Jeans equation, which is projected over the line of sight and averaged within an aperture to form the corresponding aperture velocity dispersion profile. Through dimensional analysis, the $M_\mathrm{vir} - \sigma_\mathrm{ap}$ scaling relation is devised explicitly in terms of analytical bounds for these aperture velocity dispersion profiles: allowing constraints to be placed on this relation for motivated parameter choices. We predict the $M_{200} - \sigma_\mathrm{ap}$ and $M_{500} - \sigma_\mathrm{ap}$ scaling relations, each with an uncertainty of $60.5\%$ and $56.2\%$, respectively. These halo mass estimates are found to be weakly sensitive to the halo's concentration and mass scale, and most sensitive to the size of the aperture radius in which the aperture velocity dispersion is measured, the maximum value for the halo's inner slope, and the minimum and maximum values of the velocity anisotropy. Our results show that a halo's structural and kinematic profiles impose only a minor uncertainty in estimating its mass. Consequently, spectroscopic surveys aimed at constraining the halo mass using kinematic tracers can focus on characterising other, more complex sources of uncertainty and observational systematics.  
\end{abstract}

\section{1. \quad Introduction}

In modern theories of structure formation, galaxies and galaxy clusters are predicted to be embedded within massive, non-baryonic, dark matter halos \autocite[e.g.][]{White1978,FrenkWhite1991}. Although such halos are a fundamental prediction of these theories, they have not been observed directly, instead being inferred by their gravitational influence on luminous tracers. Observational estimates for the masses of these halos are typically deduced from the observed kinematics of tracer populations such as stars, star clusters or galaxies \autocite[e.g.][]{Eke2004, Robotham2011}, from X-ray emission of hot gas within the halo's gravitational potential \autocite[e.g.][]{Vikhlinin2006, Vikhlinin2009, Babyk2023}, or from gravitational lensing \autocite[e.g.][]{Hoekstra2013}. 

The Halo Mass Function (HMF) quantifies the number density of dark matter halos in the universe per unit mass interval, and is a key prediction of any theory of
cosmological structure formation.
The general form of the HMF has been predicted analytically \autocite{PressSchechter1974, ShethTormen2002}, but accurate characterisation of its form requires $N$-body cosmological simulations. A range of functional forms have been published \autocite[see, e.g.][for a comprehensive summary]{Murray2013} with a particular focus on its form in the standard $\Lambda$-Cold Dark Matter $(\Lambda \mathrm{CDM})$ model. Meanwhile, the amplitude and shape of the HMF are sensitive to cosmological parameters \autocite[see, e.g.][]{murray2013a}, and its lower-mass end slope depends on the nature of dark matter \autocite[such as Warm Dark Matter, e.g.][]{Smith2011}.
                
When constructing the HMF, spectroscopic surveys measure galaxy spectra to estimate distance and recessional velocity, which allows one to infer the line of sight velocity distributions of large galaxies, groups and clusters of galaxies. These line of sight velocity distributions can be converted to line of sight velocity dispersions, which, when averaged within a given projected radius (i.e. aperture) relative to the centre and systemic velocity of the association, is taken to be proportional to the assumed dark matter halo mass of the galaxy, group or cluster. The parameterised relationship between halo mass and aperture velocity dispersion is typically calibrated using cosmological $N$-body simulations \autocite[e.g.][]{Eke2004, Robotham2011}, with the uncertainty in the halo mass recovered from these estimators typically of the order $0.3-0.5\,\mathrm{dex}$. These estimates are further challenged by the deficient number of group members typically available within a system, introducing a large statistical uncertainty in these calculations \autocite[see, e.g.][]{Beers1990}. 

\medskip

Recent work on measurement of the HMF using galaxies as kinematic tracers \autocite[e.g. the Galaxy And Mass Assembly (GAMA) survey,][]{GAMA2022} have highlighted the need for robust halo mass estimates that are model independent. 
This motivates the work presented in this paper: the development of a theoretical toolkit to quantify the relationship between dark matter halos with generalised, albeit spherically symmetric, mass profiles and the permitted velocity distributions of kinematic tracers embedded within them. Such a toolkit allows us to understand how uncertainties in a halo's structure (e.g. the inner slope of the density profile) and tracer kinematics (e.g. the velocity anisotropy) are likely to propagate through into observational inferences of halo masses.

To illustrate this point, note that the kinematics of tracer populations embedded within large gravitational systems that are spherically symmetric and in virial equilibrium are predictable analytically. There is a relationship between the system's mass profile, its gravitational potential, and the velocity dispersion of bound tracers on a given orbit, but the precise form is complicated by the kinds of orbits that the tracers follow (e.g. isotropic or preferentially radial). This makes separating the respective influences of the mass profile and orbital properties of the observed velocity dispersion difficult, leading to the well known `mass-anisotropy degeneracy' \autocite{BinneyMamon1982}, and represents an inherent uncertainty when seeking a relationship between halo mass and tracer kinematics. 

Quantifying the level of this uncertainty is important for a variety of reasons, but for our purposes we wish to  understand its likely impact on inferences of the HMF. We do this by adopting physically-motivated bounds on the form of the dark matter halo density profile (e.g. a range of inner slopes but a fixed outer slope) and varying the velocity anisotropy of the tracer population, in turn fixing the correspondence of the halo mass to the observed velocity dispersion within this parameter space. This approach is complementary to investigations of the structural (e.g. \cite{NFW1997}, hereafter NFW; \cite{navarro2010}) and kinematic \autocite[e.g.][]{benson2005,aung2021,bakels2021} properties obtained directly from $N$-body simulations, and allows us to explore halo mass uncertainties, simulation-independently, over a general parameter space, in a flexible and self-consistent manner. 
Moreover, by deriving a scale-free relationship between the velocity dispersion and the halo mass over such a parameter space, this halo mass scaling relation can be constrained. 

\medskip

Our theoretical tool-kit is laid out and explored in Section 2. In particular, we present an overview of predictions from $N$-body simulations and observations for the form of the dark matter halo density profile, and the analytical framework for relating a halo's structure to its kinematic profiles. Using the NFW profile as our template, we motivate a generalised halo mass profile, which we call the `ideal physical halo', and we use the virial theorem to constrain the correspondence between the halo mass and its kinematic observables. In Section 3 these kinematic profiles are derived, with an analysis of their bounds over the outlined parameter space fixing constraints on the halo mass. Scaling relations are presented Section 4, providing the central result of this study, with the dependence on the chosen parameter space outlined thereafter. We present our conclusions in Section 5. 

\section{2. \quad Theoretical background and methods}
\vspace{2mm}
\subsection{2.1. \quad The NFW profile}
We begin our theoretical background by considering 
the Navarro-Frenk-White (NFW) profile, which describes the spherically averaged mass distribution of dark matter halos
\citepalias{NFW1995, NFW1996, NFW1997}. This has been studied extensively in the literature \autocite[e.g.][]{LokasMamon2001}, and so provides us with a useful comparison when we consider our generalised profiles. 

The NFW profile, derived from fits to the ensemble average of dynamically relaxed halos in cosmological $N$-body simulations, has the density profile, $\rho(r)$, given by:
\begin{equation}\label{NFW profile definition}
    \frac{\rho (r)}{\rho _\mathrm{crit,0}} =  \frac{ \delta _\mathrm{char}}{r/r_\mathrm{s}\left(1+ {r/r_\mathrm{s}}\right)^2},
\end{equation}
where $r$ is the halocentric radius, $r_\mathrm{s}$ is the scale radius, $\rho_\mathrm{crit,0}$ is the present critical density of the universe, and $\delta_\mathrm{char}$ is the characteristic density. Equation \eqref{NFW profile definition} is observed to provide a good fit to the mass profiles of halos simulated over a large range of halo masses, cosmological parameters, and cosmological models; in this sense it is regarded as a universal profile. 


\subsubsection{The concentration parameter $c$}

Physical properties of halos are typically quoted in terms of their so-called virial parameters. These parameters describe gravitational structures that are in virial equilibrium, as the state in which its gravitational potential energy is balanced by its internal energy
(approximately; cf. \cite{colelacey1996}). 
The virial mass, $M_\mathrm{vir}$, is defined as the mass enclosed within a spherical volume of halocentric radius $r_\mathrm{vir}$, called the virial radius, whereby:
\begin{equation}\label{virial mass definition}
    M_\mathrm{vir} \equiv \frac{4}{3}\pi r_\mathrm{vir}^3 \Delta \rho_\mathrm{crit,0},
\end{equation}
such that the mean density enclosed at $r_\mathrm{vir}$ is equal to the virial overdensity parameter, $\Delta$, times the present critical density of the universe, $\rho_\mathrm{crit,0}$ \autocite[e.g.][]{white2001}. 

The spherical collapse model in an Einstein de-Sitter universe
predicts $\Delta \approx 178$, but the common convention is that $\Delta=200$, which is independent of cosmology and redshift, thus defining the virial mass $M_{200}$ and virial radius $r_{200}$. These halo parameters, $M_{200}$ and $r_{200}$, reasonably model the halo's virial mass and radius in a $\Lambda \mathrm{CDM}$ universe. An alternative, commonly used in studies of galaxy clusters, is to use $\Delta=500$; this is because observations of X-ray cluster emission are typically limited to smaller halo radii, requiring a smaller enclosed mass be predicted. This allows one to define the halo mass $M_{500}$ and the halo radius $r_{500}$. 

It is common in the literature for the NFW profile to be normalised by its virial parameters: by integrating the density profile over the volume of a sphere of halocentric radius $r_\mathrm{vir}$, and demanding the enclosed mass be equal to the virial mass. This naturally gives rise to the concentration parameter, $c$, defined as the ratio of the virial radius to the scale radius, $r_\mathrm{s}$, as:
\begin{equation}\label{concentration definition}
    c \equiv \frac{r_\mathrm{vir}}{r_\mathrm{s}}.
\end{equation}
The concentration parameter is known from $N$-body simulations to vary weakly with halo mass, in general being higher for low-mass systems (e.g. \citetalias{NFW1996}; \cite{bullock2001,ludlow2014}), with typical values in the $\Lambda \mathrm{CDM}$ cosmology of $c=5$ for cluster-scale halos, and $c=10$ for galaxy-scale halos.

Importantly, this definition of the concentration parameter depends on the choice of virial overdensity parameter, $\Delta$, that specifies the virial radius. This is because $r_{500} < r_{200}$, and by definition in Equation \eqref{concentration definition}, this implies an overdensity-dependent concentration parameter, such that $c_{500} < c_{200}$. In this paper, we refer to the concentration only as $c$, and take appropriate values when specifying different choices of $\Delta$. 

\subsubsection{The NFW profile: scale-free form}

We introduce a dimensionless radial scale, $s$, defined as:
\begin{equation}\label{dimensionless radius definition}
    s \equiv \frac{r}{r_\mathrm{vir}},
\end{equation}
where $r$ is the halocentric radius. This allows us to rewrite Equation \eqref{NFW profile definition} in the scale-free form:
\begin{equation}\label{scale-free NFW profile}
    \frac{\rho (s, c)}{\Delta \rho _\mathrm{crit,0}} =  \frac{g(c)}{3s(1+cs)^2},
\end{equation}
where $c$ is the NFW concentration and $g(c)$ is called the NFW concentration function, defined as:
\begin{equation}\label{NFW concentration function}
    g(c) = \frac{c^2}{\ln(1+c) - \frac{c}{(1+c)}}.
\end{equation}

\subsubsection{The NFW gravitational potential}
The gravitational potential, $\Phi (r)$, of a spherically symmetric halo of density, $\rho (r)$, is given by:
\begin{equation}\label{gravitational potential}
    \Phi(r) = -4\pi G \left[\frac{1}{r}\int_0^r r^{\prime 2} \rho(r^\prime)\mathrm{d}r^\prime + \int_r^\infty r^\prime \rho (r^\prime)\mathrm{d}r^\prime \right],
\end{equation}
where $G$ is Newton's gravitational constant, and $r$ is the halocentric radius. In scale-free form, normalised in terms of virial parameters, the gravitational potential can be expressed as:
\begin{equation}\label{scale-free gravitational potential}
    \frac{\Phi(s)}{v_\mathrm{vir}^2} = -3 \left[\frac{1}{s}\int_0^s s^{\prime}{}^2 \frac{\rho(s^\prime)}{\Delta \rho_\mathrm{crit,0}} \mathrm{d}s^\prime + \int_s^\infty s^\prime \frac{\rho (s^\prime)}{\Delta \rho_\mathrm{crit,0}} \mathrm{d}s^\prime \right],
\end{equation}
in the ratio to the square of the virial circular velocity, $v_\mathrm{vir}$, defined as the circular velocity of a particle orbiting a gravitational mass $M_\mathrm{vir}$, at a radius $r_\mathrm{vir}$, as:
\begin{equation}\label{virial velocity definition}
    v_\mathrm{vir}^2 \equiv \frac{GM_\mathrm{vir}}{r_\mathrm{vir}}.
\end{equation}
Taking the form of the NFW profile as in Equation \eqref{scale-free NFW profile}, the associated gravitational potential is:
\begin{equation}\label{NFW gravitational potential}
    \frac{\Phi(s, c)}{v_\mathrm{vir}^2}=-\frac{g(c)}{c^2}\frac{\ln(1+cs)}{s}.
\end{equation}

\subsection{2.2. \quad Constructing a generalised halo profile}

We now consider the case of the generalised halo profile. We assume spherical symmetry and a density profile with a logarithmic density slope (hereafter, for brevity, slope) that is shallower at small radius and steeper at larger radius. The NFW profile is one specific example of this generalisation.

There is reasonable consensus that the outer slope of virialised dark matter halos varies as $\rho (r) \sim r^{-3}$, as found in numerical simulations (e.g. \citetalias{NFW1996}), consistent with observations of galaxy kinematics \autocite[e.g.][]{Prada2003} and theoretically expected for typical halo mass assembly histories \autocite[e.g.][]{Lu2006,ludlow2013}. However, comparison of theoretical predictions and observational limits on the inner slope is complicated, because of the complex coupling between dark matter and baryons during galaxy assembly, such that the inner density of dark matter halos may be sensitive to, e.g. its mass accretion history or episodic stellar feedback, which may have a complex dependence on halo mass \autocite[e.g.][]{DiCintio2014,Chan2015,tollet2016}.
For this reason, we focus on spherically symmetric halos with outer slope $-3$, and inner slope $-\alpha$, with $\rho (r) \sim r^{-\alpha}$ at small halocentric radii. We refer to these generalised NFW halos as `ideal physical halos'. 

\subsubsection{The dark matter halo inner slope $\alpha$}

The NFW profile is the $\alpha=1$ member of the class of ideal physical halos, with a divergent density profile, i.e. a `cusp', in the central region. If instead the density profile were to flatten to some constant value in the central region, such that $\alpha \simeq 0$, this would instead be referred to as a `core'. 

Numerical simulations consistently predict halo cusps, while observations indicate that the halos of dark matter dominated galaxies are cored \autocite[i.e., $\alpha=0$; e.g.][]{Moore1994,deBlok2003},
a tension known as the `core-cusp problem'. The strength of the predicted inner cusp, as measured by $\alpha$, could be as steep as $\alpha \simeq 1.5$;
this has been predicted in simulations of initial proto-halo structures undergoing gravitational collapse \autocite[][]{OgiyaHahn2017}.
Motivated by these results, we investigate inner slopes sensibly bounded by the range of values $\alpha \in [0, 1.5]$: incorporating both cores and cusps, applicable to halos over a wide-range of masses, formation histories and feedback physics, and agnostic to the tension between simulated and observed halo structures. 

\subsubsection{The ideal physical halo profile}
Analogous to the virialised, scale-free NFW profile, we can construct a corresponding density profile to describe the ideal physical halos, as a function of the dimensionless radius, $s$, and parameterised by its concentration, $c$, and inner slope, $\alpha$, such that:
\begin{equation}\label{ideal physical halo profile}
    \frac{\rho(s, c, \alpha)}{\Delta \rho_\mathrm{crit,0}} = \frac{u(c, \alpha)}{3s^\alpha (1+cs)^{3 - \alpha }},
\end{equation}
with $u(c, \alpha)$ representing a generalised concentration function, defined by the integral:
\begin{equation}\label{ideal physical halo concentration function}
    u(c, \alpha) \equiv \left[\int _0^1\frac{s^{2-\alpha }\mathrm{d}s}{(1+cs)^{3 - \alpha }}\right]^{-1}.
\end{equation}
In this form, the NFW profile from Equation \eqref{scale-free NFW profile} is recovered by setting $\alpha = 1$, with the associated NFW concentration function recovered as $g(c) \equiv u(c, \alpha = 1)$. 

\subsubsection{The ideal physical halo gravitational potential}\label{2.2.3}

The scale-free gravitational potential of an ideal physical halo is then given by the profile
\begin{equation}\label{ideal physical halo gravitational potential}
    \frac{\Phi(s, c, \alpha)}{v^2_\mathrm{vir}}=-{u(c, \alpha)}\left[\frac{1}{s}\int_0^s \frac{s^\prime {}^{2-\alpha}\mathrm{d}s^\prime  }{(1+cs^\prime )^{3 - 
    \alpha}} + \int_s^\infty \frac{s^\prime {}^{1-\alpha }\mathrm{d}s^\prime  }{(1+cs^\prime )^{3 - \alpha}} \right].
\end{equation}

\subsection{2.3. \quad Kinematic profiles of dark matter halos}\label{Section 2.3}


Here we consider the kinematics of a non-dissipational tracer population (e.g. stars, galaxies) in the gravitational potential of a dark matter halo.

\subsubsection{The velocity dispersion of tracer populations}
The components of the orbital velocity of a tracer population will vary depending on its radius within its host dark matter halo.
The standard deviation of these velocities over the tracer population is known as the velocity dispersion. Typically, the velocity anisotropy parameter, $\beta(r)$, \textcolor{black}{as introduced in \cite{Binney1980}}, is used to characterise the relation between the angular and radial components of the velocity dispersion; this measure is typically defined as:
\begin{equation}\label{velocity anisotropy definition}
    \beta(r) \equiv 1 - \frac{\sigma _\theta^2 (r) }{\sigma _\mathrm{r} ^2 (r)},
\end{equation}
as a function of halocentric radius $r$, and where $\sigma _\theta(r)$ and $\sigma_\mathrm{r}(r)$ are the angular and radial velocity dispersions, respectively. 

\subsubsection{The Jeans equation}

The relationship between the radial velocity dispersion, $\sigma_\mathrm{r}(r)$, the density, $\rho(r)$, and the gravitational potential, $\Phi(r)$, of a spherically symmetric mass distribution is described by the Jeans equation:
\begin{equation}\label{Jeans equation}
    \frac{\mathrm{d}}{\mathrm{d}r}\left[\rho(r)\sigma_\mathrm{r}^2(r)\right]+\frac{2\beta(r)}{r}\rho(r)\sigma_\mathrm{r}^2(r)=-\rho(r)\frac{\mathrm{d}\Phi(r)}{\mathrm{d}r}.
\end{equation}
This differential equation encodes a `mass-anisotropy degeneracy', as the velocity dispersion depends on a coupling of both the anisotropy and the underlying density and gravitational potential of the system, which can be challenging to disentangle in kinematic observations. 

In the simple case of a constant anisotropy, $\beta (r) = \beta$, the general solution to the Jeans equation is \textcolor{black}{\autocite{Binney1980, BinneyTremaine2008}:}
\begin{equation}\label{radial velocity dispersion}
    \sigma_\mathrm{r}^2(r, \beta) = \frac{r^{-2\beta}}{\rho(r)}\int_r ^\infty r^\prime {}^{2\beta} \rho (r^\prime)\frac{\mathrm{d}\Phi(r^\prime) }{\mathrm{d}r^\prime }\mathrm{d}s^\prime.
\end{equation}
In our virial, scale-free framework, this general solution is given by the profile:
\begin{equation}\label{scale-free radial velocity dispersion}
    \frac{\sigma_\mathrm{r}^2(s, \beta)}{v_\mathrm{vir}^2} =s^{-2\beta} \frac{\Delta \rho_\mathrm{crit,0}}{\rho(s)}\int_s ^\infty s^\prime {}^{2\beta} \frac{\rho (s^\prime)}{\Delta \rho_\mathrm{crit,0}} \frac{\mathrm{d}}{\mathrm{d}s^\prime }\left[\frac{\Phi(s^\prime)}{v_\mathrm{vir}^2}\right] \mathrm{d}s^\prime.
\end{equation}

\subsubsection{The velocity anisotropy parameter $\beta$}

To model the velocity dispersion of a dark matter halo, reasonable physical values for the anisotropy parameter, $\beta$, need to be assumed. Theoretically, the anisotropy parameter can vary between purely radial orbits, of $\beta = 1$ (when $\sigma_\theta = 0$), isotropic orbits, of $\beta = 0$ (when $\sigma_\theta$ = $\sigma_\mathrm{r}$) and purely circular orbits, when $\beta \to -\infty$ (when $\sigma_\mathrm{r} = 0$).

Numerical simulations of galaxy clusters have consistently predicted isotropic orbits in the central region of a halo \autocite[e.g.][]{Debattista2008}, with its spherically-averaged value increasing radially from the centre of a halo up to values of $\beta \simeq 0.25-35$ near $r_{200}$ \autocite[e.g.][]{Thomas1998, Lemze2012}. Some observational surveys have measured the velocity anisotropy within galaxy clusters by calibrating X-ray observables to a predicted universal relation between the dark matter and gas temperature recovered from simulations \autocite[e.g.][]{Host2009}: yielding an average value of $\beta \simeq 0.3$ in the central region that radially increases toward $\beta \simeq 0.5$ at about $20-30\%$ of $r_{200}$, where this calibration diminishes. 

As these studies consistently predict a spherically-averaged anisotropy within $0\lesssim \beta \lesssim 0.5$ between the halo's centre and $r_{200}$, we will take this bound $\beta \in [0, 0.5]$ to model the kinematics of the ideal physical halos. 

\subsubsection{The projected surface mass density profile}

When observers make contact with the velocity dispersion of tracer populations, information is projected along the line of sight direction. The projection of the halo density profile is the surface mass density profile, $\Sigma (R)$, as a function of the projected radius, $R$: the projection of the halocentric radius, $r$, along the line of sight direction. For a spherically symmetric mass distribution of density profile $\rho(r)$, its surface mass density is given by the projection:
\begin{equation}\label{surface mass density}
    \Sigma(R) = 2\int_R^\infty \frac{r\rho(r)\mathrm{d}r}{\sqrt{r^2-R^2}}.
\end{equation}
Anticipating a scale-free description, we introduce a dimensionless projected radius, $S$, as the ratio of the projected radius to the virial radius, whereby:
\begin{equation}\label{dimensionless projected radius definition}
    S \equiv \frac{R}{r_\mathrm{vir}},
\end{equation}
such that the surface mass density can be expressed by the profile:
\begin{equation}\label{scale-free surface density}
    \frac{\Sigma(S)}{\Sigma_\mathrm{vir}} = \frac{3}{2}\int_{S}^\infty \frac{\rho(s)}{\Delta \rho_\mathrm{crit,0}}\frac{s\mathrm{d}s}{\sqrt{s^2-S^2}}.
\end{equation}
In the above profile, we have scaled the surface density by the virial surface density, $\Sigma_\mathrm{vir}$, defined as the mean surface density corresponding to a virial mass, $M_\mathrm{vir}$, when projected within a cylinder of projected radius $r_\mathrm{vir}$, such that:
\begin{equation}\label{virial surface density definition}
    \Sigma_\mathrm{vir} \equiv \frac{M_\mathrm{vir}}{\pi r_\mathrm{vir}^2}.
\end{equation}

\subsubsection{The line of sight velocity dispersion}

To make contact with the velocity dispersion of a galaxy or cluster, observers measure the projection of the radial velocity dispersion along the line of sight, known as the line of sight velocity dispersion, $\sigma_\mathrm{los}(R)$, as observed at some projected radius, $R$. For a spherical non-rotating gravitational system, the line of sight velocity dispersion of a system with constant anisotropy, $\beta$, can be recovered from the integral expression \autocite{BinneyMamon1982}:
\begin{equation}\label{los velocity dispersion}
    \sigma _\mathrm{los} ^2 (R, \beta ) = \frac{2}{\Sigma (R)} \int _R^\infty  \frac{\left(1 -  \frac{\beta R^2}{r^2}\right) \rho (r) \sigma _\mathrm{r} ^2 (r, \beta) r \mathrm{d}r}{\sqrt{r^2 - R^2}},
\end{equation}
where $\sigma_\mathrm{r}(r, \beta)$ is the constant-anisotropy radial velocity dispersion, and $\rho(r)$ and $\Sigma (R)$ are the density and surface mass density profiles, respectively. In terms of the dimensionless variables, $s$ and $S$, in a scale-free formulation, this observable takes the form:
\begin{equation}\label{scale-free los velocity dispersion}
\begin{aligned}
    \frac{\sigma_\mathrm{los} ^2(S, \beta )}{v_\mathrm{vir}^2} &= \frac{3}{2}\frac{\Sigma_\mathrm{vir}}{\Sigma(S)} \int_{S}^\infty  \frac{\left(1-\frac{\beta S^2}{s^2}\right) \frac{\rho(s)}{\Delta \rho_\mathrm{crit,0}} \frac{\sigma_\mathrm{r}^2(s,\beta)}{v_\mathrm{vir}^2} s\mathrm{d}s }{\sqrt{s^2-S^2}}.
\end{aligned}
\end{equation}
Of note, this integral expression for the line of sight velocity dispersion can be simplified in the case of isotropic, $\beta=0$ orbits, reducing to the form:
\begin{equation}\label{scale-free isotropic los velocity dispersion}
    \frac{\sigma _\mathrm{los} ^2 (S, \beta=0)}{v_\mathrm{vir}^2} = \frac{3}{2} \frac{\Sigma_\mathrm{vir} }{\Sigma (S)} \int _{S}^\infty \sqrt{s^2 - S^2}\frac{\rho(s)}{\Delta \rho_\mathrm{crit,0}} \frac{\mathrm{d}}{\mathrm{d}s}\left[\frac{\Phi(s)}{v_\mathrm{vir}^2}\right]\mathrm{d}s,
\end{equation}
which provides a convenient simplification when numerically integrating these profiles. 

\subsubsection{The aperture velocity dispersion}

Due to the poor statistics of group members usually recovered in observations, the line of sight velocity dispersion of all group members is typically averaged within a fixed projected radius, called as the aperture radius, $R_\mathrm{ap}$. This aperture-averaged velocity dispersion is known as the aperture velocity dispersion, $\sigma_\mathrm{ap}(<R_\mathrm{ap})$. Analytically, this observable can be derived from the integral expression \autocite{LokasMamon2001}: 
\begin{equation}\label{aperture velocity dispersion}
    \sigma_\mathrm{ap}^2(<R_\mathrm{ap}, \beta) = \frac{\int _0 ^{R_\mathrm{ap}} \Sigma (R) \sigma_\mathrm{los} ^2 (R, \beta) R \mathrm{d}R }{\int _0 ^{R_\mathrm{ap}} \Sigma (R) R \mathrm{d}R},
\end{equation}
in terms of some constant anisotropy, $\beta$, and where $\sigma_\mathrm{los} (R, \beta)$ is the constant-anisotropy line of sight velocity dispersion, and $\Sigma(R)$ is the surface mass density profile. Finally, we can express this observable in our scale-free formalism, in terms of a dimensionless aperture radius, $S_\mathrm{ap}$,
\begin{equation}\label{scale-free aperture velocity dispersion}
    \frac{\sigma_\mathrm{ap}^2(<S_\mathrm{ap}, \beta)}{v_\mathrm{vir}^2} = \frac{\int _0 ^{S_\mathrm{ap}} \frac{\Sigma (S)}{\Sigma_\mathrm{vir}} \frac{\sigma_\mathrm{los} ^2 (S, \beta)}{v_\mathrm{vir}^2} S \mathrm{d}S }{\int _0 ^{S_\mathrm{ap}} \frac{\Sigma (S)}{\Sigma_\mathrm{vir}}  S \mathrm{d}S}.
\end{equation}

\subsection{2.4. \quad Predictions for the $M_\mathrm{halo} - \sigma_\mathrm{ap}$ scaling relation}

With these scale-free kinematic profiles for the ideal physical halos, we now seek to establish a correspondence between the aperture velocity dispersion, $\sigma_\mathrm{ap}$, and the halo mass, $M_\mathrm{halo}$, in the form of a scaling relation. 

\subsubsection{Dimensional analysis}

To derive this scaling relation, we use the method of dimensional analysis: allowing the form of the relationship between different physical quantities to be predicted from dimensional requirements of their base units. For the desired scaling relation, we seek to establish a correspondence between the halo's mass, $M_\mathrm{halo}$ (measured in $\mathrm{M}_\odot$), and its velocity dispersion, $\sigma_\mathrm{ap}$ (measured in $\mathrm{km \, s^{-1}}$), by positing some combination of physical constants to ensure dimensionality, and by introducing a dimensionless prefactor, $\mathcal{A}$. In this case, we can reasonably assume the correspondence depends on the physical constants: $G$ (measured in $\mathrm{Mpc \, km^2\,s^{-2} \, \mathrm{M}_\odot ^{-1}}$), and $\rho_\mathrm{crit,0}$ (measured in $\mathrm{M_\odot \, Mpc^{-3}}$), such that dimensional requirements entail the form:
\begin{equation}\label{dimensional analysis}
    M_\mathrm{halo} = \mathcal{A} \cdot \sqrt{\frac{1}{\rho_\mathrm{crit,0} G^3}} \cdot \sigma_\mathrm{ap}^3,
\end{equation}
with quantitative constraints on this relation dependent on a prediction for this dimensionless prefactor, $\mathcal{A}$. 

To constrain $\mathcal{A}$, we can use the virial theorem to relate the halo's virial mass to its virial circular velocity, by utilising the definitions in Equations \eqref{virial mass definition} and \eqref{virial velocity definition}, whereby:
\begin{equation}\label{virial mass - velocity relation}
    M_\mathrm{vir} = \sqrt{\frac{3}{4\pi \Delta \rho_\mathrm{crit,0} G^3}} \cdot v_\mathrm{vir} ^3,
\end{equation}
which takes on the same dimensional form as Equation \eqref{dimensional analysis} above. By taking the virial mass, $M_\mathrm{vir}$, in either approximation $M_{200}$ or $M_{500}$, as an approximation to the halo mass, $M_\mathrm{halo}$, we can use this correspondence to recover the scaling relation:
\begin{equation}\label{virial mass bound}
    M_\mathrm{vir} = \sqrt{\frac{3}{4\pi \Delta \rho_\mathrm{crit,0} G^3}} \cdot \left[\frac{\sigma_\mathrm{ap}}{\xi}\right]^3,
\end{equation}
composed in terms of the aperture velocity dispersion, $\sigma_\mathrm{ap}$, by introduction of the dimensionless parameter, $\xi$, defined as:
\begin{equation}\label{xi definition}
    \xi \equiv \frac{\sigma_\mathrm{ap}(<R_\mathrm{ap})}{v_\mathrm{vir}}.
\end{equation}
This dimensionless parameter, $\xi$, is the scale-free form of the aperture velocity dispersion, measured within a specified aperture radius, $R_\mathrm{ap}$, and so can be determined and constrained analytically from Equation \eqref{scale-free aperture velocity dispersion} when modelling the ideal physical halos.

\subsubsection{Constraining the scaling relation}

When determining this parameter $\xi$ continuously over the parameter space of ideal physical halos --- for halos with inner slope $\alpha \in [0, 1.5]$ and velocity anisotropy $\beta \in [0, 0.5]$ --- its value will be bounded in some region, $\xi \in [\xi_\mathrm{min}, \xi_\mathrm{max}]$, when measured within some given aperture radius, $R_\mathrm{ap}$. 

From this analytic modelling, the bounds predicted for $\xi$ will allow the scaling relationship $M_\mathrm{vir} - \sigma_\mathrm{ap}$ in Equation \eqref{virial mass bound} to be constrained, as bounded within a corresponding minimum and maximum proportionality. This technique for constraining the scaling relation quantifies the form predicted by dimensional analysis, and in doing so circumvents the mass-anisotropy degeneracy that arises in the Jeans equation within the prescribed bounds in halo parameters.

Taking the values of the physical constants in Equation \eqref{virial mass bound}, the halo mass scaling relation reduces to the correspondence:
\begin{equation}\label{virial mass bound - evaluated}
    \frac{M_\mathrm{vir}}{\mathrm{M}_\odot} = \frac{3.288 \times 10^6}{\xi ^3 \Delta ^{1/2} h} \cdot \left[\frac{\sigma_\mathrm{ap}}{\mathrm{km \, s^{-1}} }\right]^3,
\end{equation}
where $h$ is the Hubble parameter, $h\equiv H_0/100\mathrm{km\,s^{-1} Mpc^{-1} }$, for $H_0$ the Hubble constant. The Hubble parameter has been measured to high precision from the Cosmic Microwave Background, within an error of one percent or less. In this study, we take the Planck result $h= 0.6751$ \autocite{Planck2016}, neglecting its uncertainty as this error will be tiny compared to the anticipated bounds in the scaling relation. 

In this $M_\mathrm{vir} - \sigma_\mathrm{ap}$ correspondence, the overdensity, $\Delta$, must be chosen to specify the virial mass approximation. Taking the convention of $\Delta=200$ in Equation \eqref{virial mass bound - evaluated}, the $M_{200} - \sigma_\mathrm{ap}$ scaling relation is predicted in the form:
\begin{equation}\label{M200 mass bound}
    \frac{M_{200}}{\mathrm{M}_\odot} = \frac{3.444 \times 10^5}{\xi ^3} \cdot \left[\frac{\sigma_\mathrm{ap}}{\mathrm{km \, s^{-1}} }\right]^3,
\end{equation}
and similarly, when taking $\Delta=500$, the $M_{500} - \sigma_\mathrm{ap}$ scaling relation is predicted in the form:
\begin{equation}\label{M500 mass bound}
    \frac{M_{500}}{\mathrm{M}_\odot} = \frac{2.178 \times 10^5}{\xi ^3} \cdot \left[\frac{\sigma_\mathrm{ap}}{\mathrm{km \, s^{-1}} }\right]^3.
\end{equation}
In this study, we will predict these scaling relations for both the $M_{200}$ and $M_{500}$ virial mass approximations. Thus, by deriving values for $\xi_\mathrm{min}$ and $\xi_\mathrm{max}$ within our scale-free framework of idealised halos, corresponding to these two virial conventions, the halo mass scaling relations can be analytically predicted. 

\subsection{2.5. \quad Regimes to bound the $M_\mathrm{vir} - \sigma_\mathrm{ap}$ scaling relation}

When deriving these bounds in $\xi$ over the parameter space of the ideal physical halos, for $\alpha \in [0, 1.5]$ and $\beta \in [0, 0.5]$, two additional parameters must be fixed: the halo concentration, $c$, and the aperture radius, $R_\mathrm{ap}$. These two parameters, in our scale-free formalism, depend on the choice in overdensity, $\Delta$. As such, these parameters must be evaluated at appropriate values when $\Delta=200$ and $\Delta=500$, to separately bound each of the $M_{200} - \sigma_\mathrm{ap}$ and $M_{500} - \sigma_\mathrm{ap}$ scaling relations. 

Additionally, in this study we will consider these scaling relations separately at cluster-scale halo masses and galaxy-scale halo masses, to take into account the mass dependence of the concentration parameter, $c$. As such, in this study we will consider four distinct regimes in which to bound $\xi$, allowing us to predict four scaling relations: each requiring a fixed choice for the scale-dependent parameters of $c$ and $R_\mathrm{ap}$. 

\subsubsection{Fixing the concentration parameter $c$ in each regime}

For the concentration $c$, we will assume the typical $\Lambda \mathrm{CDM}$ values when $\Delta=200$: fixing its value to $c=5$ for cluster-scale halos and $c=10$ for galaxy-scale halos. When $\Delta=500$, we will assume to first order that $r_{500}/r_{200} \simeq 0.5$\footnote{This value $r_{500}/r_{200}$ will depend on the specific form of the halo's density profile. As both $r_{200}$ and $r_{500}$ are both assumed to be approximations to the virial radius, this conversion is not of strong importance, with any modification in their ratio being purely quantitative.}, reducing the concentration values to $c=2.5$ for cluster-scale halos and $c=5$ for galaxy-scale halos. 

\subsubsection{Fixing the aperture radius $R_\mathrm{ap}$ in each regime}

For the aperture radius, $R_\mathrm{ap}$, a range in values is more appropriate than setting a fixed value: as observational surveys that measure the aperture velocity dispersion are usually limited in the number of tracer populations available in a given system, and the aperture radius chosen to calculate the velocity dispersion of the observable tracers will vary accordingly. 

In our scale-free formalism, this aperture radius appears in dimensionless form $S_\mathrm{ap}\equiv R_\mathrm{ap}/r_\mathrm{vir}$, and so any range in values must be specified in units of $r_{200}$ (when $\Delta=200$) or $r_{500}$ (when $\Delta=500$). For this reason, we will take the aperture radius between $R_\mathrm{ap} = 0.1 r_{200}$ and $R_\mathrm{ap} = r_{200}$, which in units of $r_{500}$ we can equate as approximately $R_\mathrm{ap} = 0.2 r_{500}$ and $R_\mathrm{ap} = 2 r_{500}$, when assuming again to first order a factor of half between these radii. 

\subsubsection{The four regimes to bound the $M_\mathrm{vir} - \sigma_\mathrm{ap}$ scaling relation}

Thus, when evaluating $\xi$ continuously over the parameter space $\alpha \in [0, 1.5]$ and $\beta \in [0, 0.5]$ for the ideal physical halos, the additional scale-dependent parameters $R_\mathrm{ap}$ and $c$ must be evaluated in four distinct regimes, with the constraints on $\xi$ in each regime constraining one of four distinct scaling relations:

\begin{enumerate}
    \item The $M_{200} - \sigma_\mathrm{ap}$ scaling relation for cluster-scale halos: \\
    with parameters: $R_\mathrm{ap} \in [0.1, 1] r_{200}, \quad c=5$.

    \item The $M_{200} - \sigma_\mathrm{ap}$ scaling relation for galaxy-scale halos: \\
    with parameters: $R_\mathrm{ap} \in [0.1, 1] r_{200}, \quad c=10$.

    \item The $M_{500} - \sigma_\mathrm{ap}$ scaling relation for cluster-scale halos: \\
    with parameters: $R_\mathrm{ap} \in [0.2, 2] r_{500}, \quad c=2.5$.

    \item The $M_{500} - \sigma_\mathrm{ap}$ scaling relation for galaxy-scale halos: \\
    with parameters: $R_\mathrm{ap} \in [0.2, 2] r_{500}, \quad c=5$.
\end{enumerate}

\begin{figure*}[h!]
    \centering
    \includegraphics[width=\textwidth]{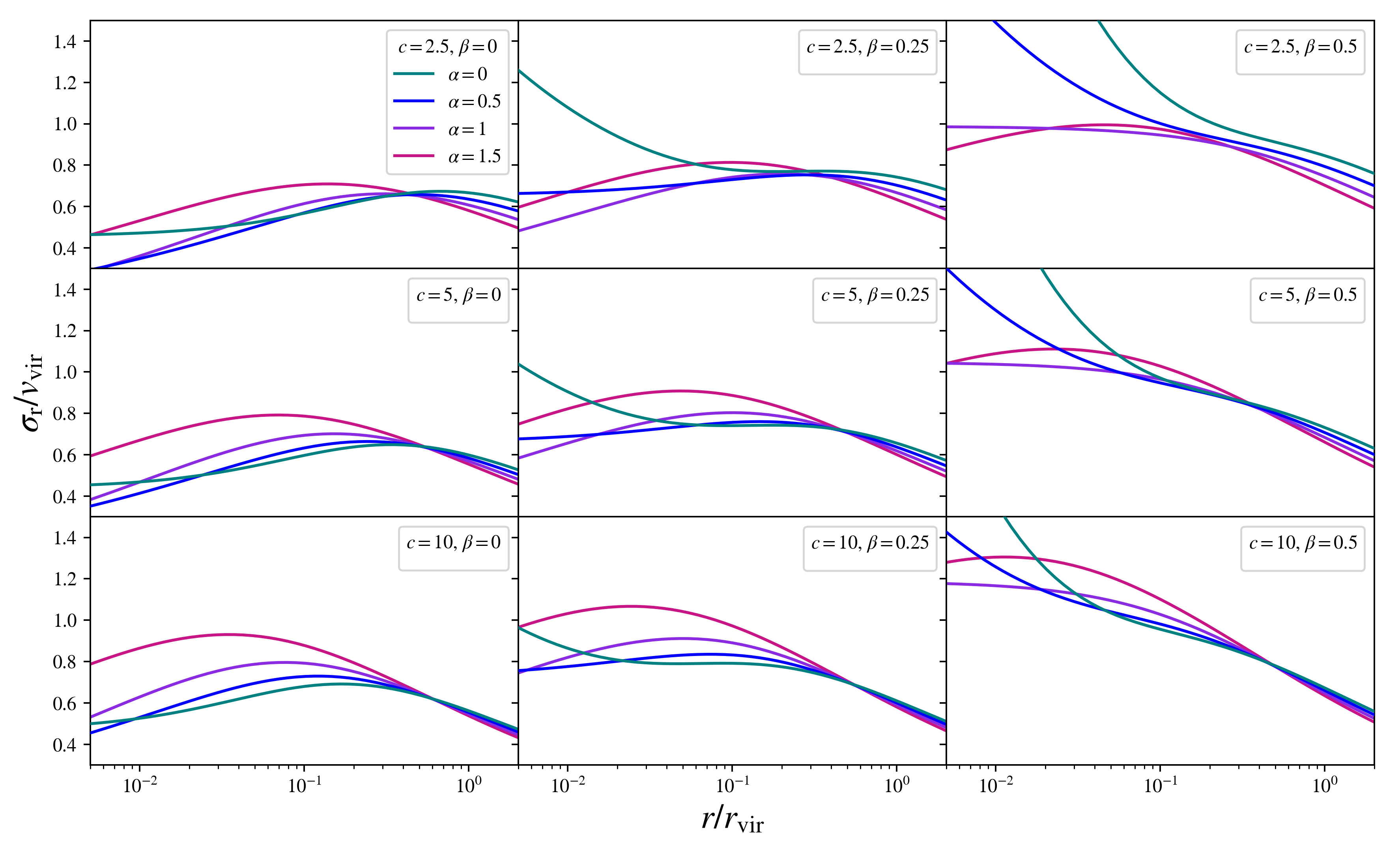}
    \caption{The radial velocity dispersion profiles for the ideal physical halos, in scale-free form $\sigma_\mathrm{r} /v_\mathrm{vir}$, traced over the scaled halocentric radius, $r/r_\mathrm{vir}$. Each row varies the halo concentration, $c$, and each column varies the velocity anisotropy, $\beta$. Within each box, each colour varies the halo inner slope, $\alpha$.}
    \label{Fig - radial velocity dispersion profiles}
\end{figure*}

\begin{figure*}[h!]
    \centering
    \includegraphics[width=\textwidth]{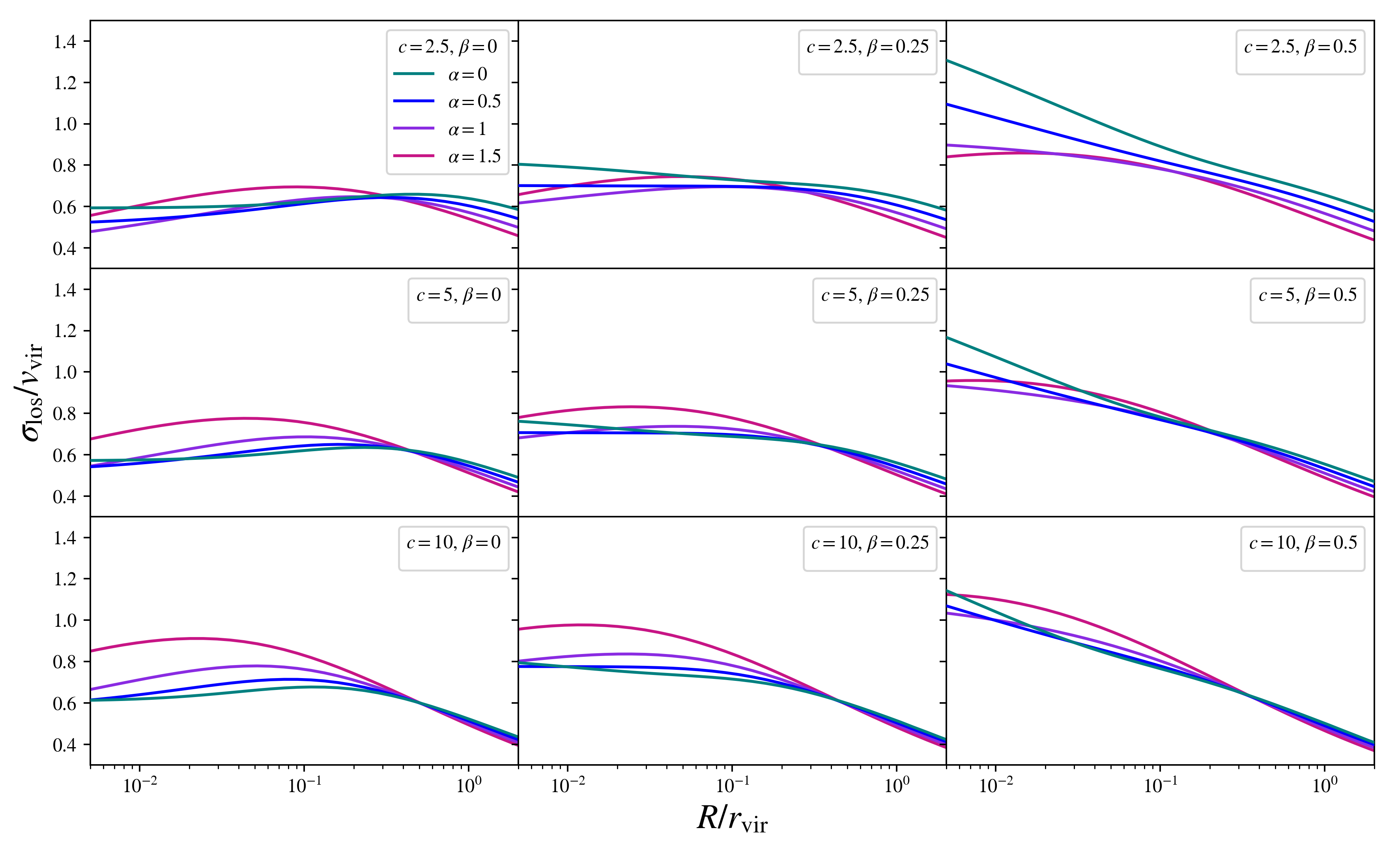}
    \caption{The line of sight velocity dispersion profiles for the ideal physical halos, in scale-free form $\sigma_\mathrm{los} /v_\mathrm{vir}$, traced over the scaled projected radius, $R/r_\mathrm{vir}$. Each row varies the halo concentration, $c$, and each column varies the velocity anisotropy, $\beta$. Within each box, each colour varies the halo inner slope, $\alpha$.}
    \label{Fig - los velocity dispersion profiles}
\end{figure*}

\begin{figure*}[h!]
    \centering
    \includegraphics[width=\textwidth]{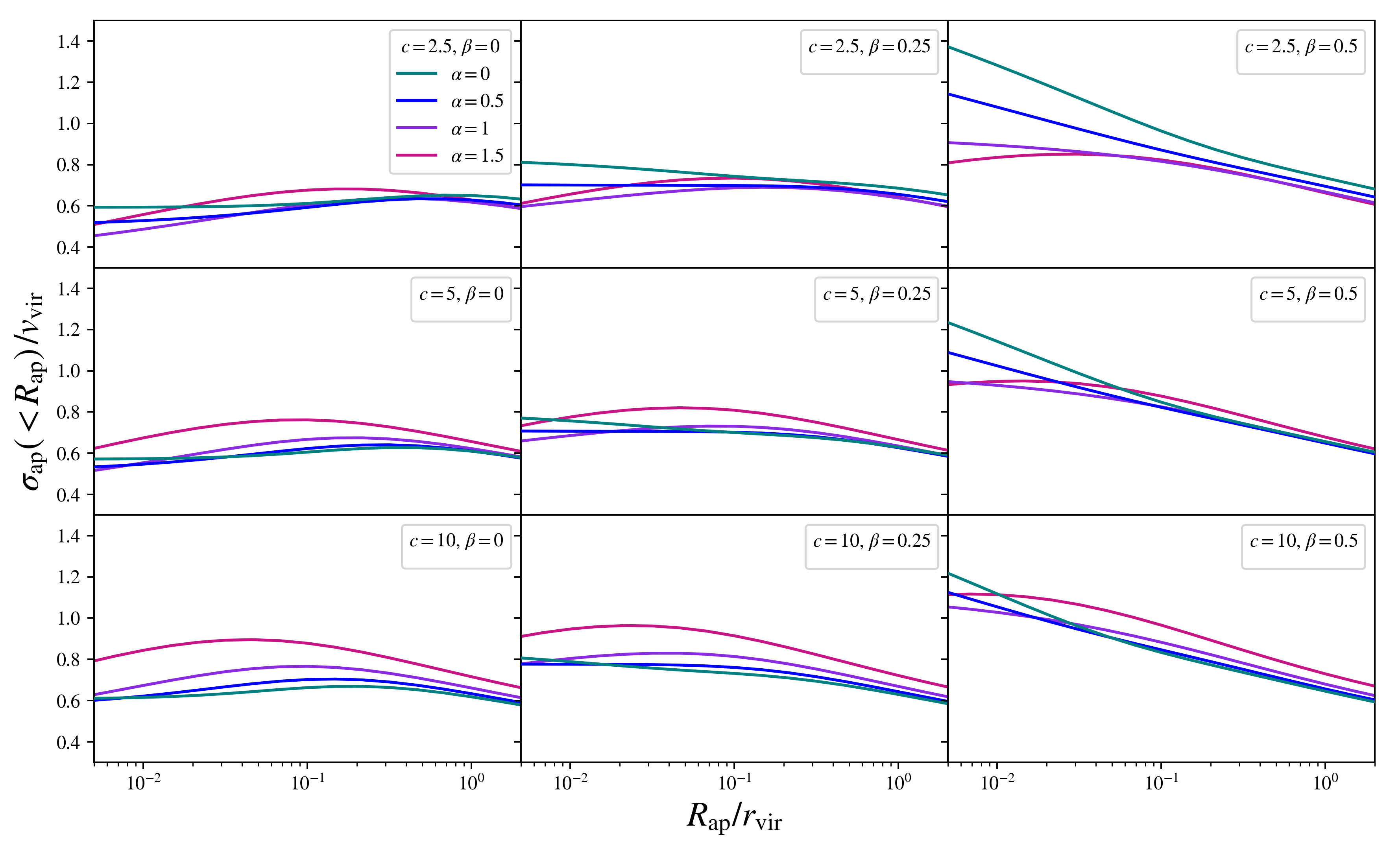}
    \caption{The aperture velocity dispersion profiles for the ideal physical halos, in scale-free form $\xi \equiv \sigma_\mathrm{ap} (<R_\mathrm{ap})/v_\mathrm{vir}$, traced over the scaled aperture radius, $R_\mathrm{ap}/r_\mathrm{vir}$. Each row varies the halo concentration, $c$, and each column varies the velocity anisotropy, $\beta$. Within each box, each colour varies the halo inner slope, $\alpha$.}
    \label{Fig - aperture velocity dispersion profiles}
\end{figure*}

\renewcommand{\arraystretch}{1.2}
\begin{table*}
\begin{tabular}{ |p{2.8cm}|p{2.2cm}|p{1.7cm}||p{1.6cm}|p{2cm}| }
\hline
Overdensity parameter & Aperture radius & Concentration  & \multicolumn{2}{|c|}{ Bounds in $\xi\equiv \sigma_\mathrm{ap}(<R_\mathrm{ap})/v_\mathrm{vir}$ } \\ [1ex]
\hline\hline
\rowcolor{white} \multirow{2}*{\vspace{1mm} $\Delta = 200$}   & \multirow{2}*{\vspace{1mm} $R_\mathrm{ap} \in [0.1, 1] \, r_\mathrm{200}$}   & $c=5$ &   $\xi_\mathrm{min} = 0.605$ & $\xi_\mathrm{max} =0.877$ \\ [1ex] 
\rowcolor{white}  & & $c=10$ &  $\xi_\mathrm{min} =0.618$  & $\xi_\mathrm{max} =0.966$ \\ [1ex]
 \hline
\rowcolor{white} \multirow{2}*{\vspace{1mm} $\Delta = 500$}   &\multirow{2}*{\vspace{1mm}$R_\mathrm{ap} \in [0.2, 2] \, r_\mathrm{500}$}   & $c=2.5$ &   $\xi_\mathrm{min} =0.584$ & $\xi_\mathrm{max} =0.880$ \\ [1ex]
\rowcolor{white}  & & $c=5$ &  $\xi_\mathrm{min} =0.576$  & $\xi_\mathrm{max} =0.822$  \\ [1ex]
\hline
\end{tabular}
\caption{The constraints placed on $\xi \equiv \sigma_\mathrm{ap} (<R_\mathrm{ap})/v_\mathrm{vir}$ over the parameter space of the ideal physical halos, in the four outlined regimes, corresponding to two conventions in the overdensity, $\Delta=200$ and $\Delta=500$, and two halo mass scales, galaxy and cluster masses, set by the concentration values.}
\label{Table - xi bounds}
\end{table*}

\begin{figure*}[h!]
    \centering
    \includegraphics[width=\textwidth]{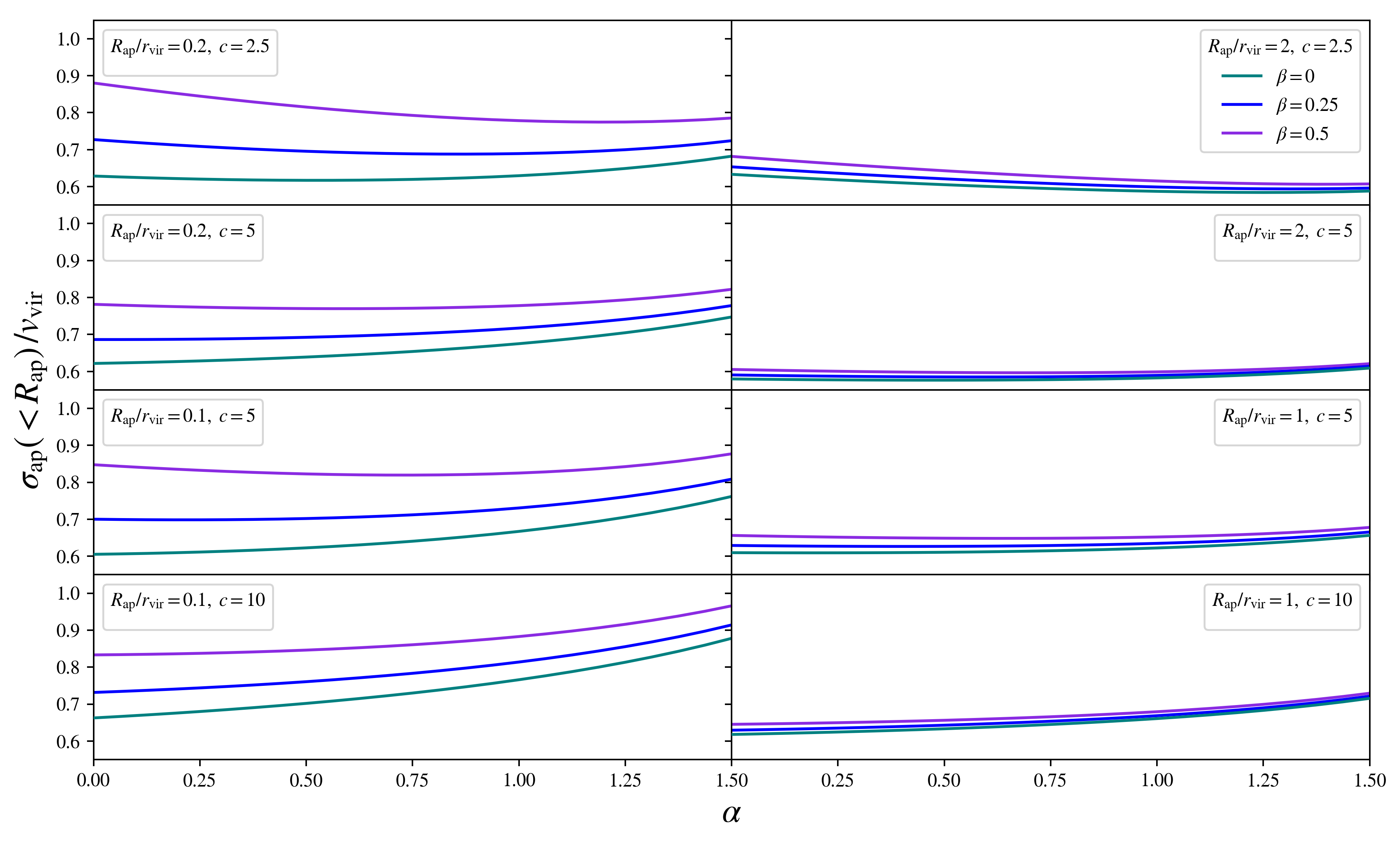}
    \caption{The aperture velocity dispersion profiles for the ideal physical halos, in scale-free form $\xi  \equiv \sigma_\mathrm{ap}(<R_\mathrm{ap})/v_\mathrm{vir}$, evaluated at fixed aperture radii and traced over halo inner slopes, $\alpha$. Each row fixes the halo concentration, $c$, and evaluates the profile at a minimum (left column) and maximum (right column) value for the aperture radius, in scale-free form $R_\mathrm{ap}/r_\mathrm{vir}$. These fixed parameters correspond to particular choices in the overdensity, $\Delta$, and the halo mass scale: corresponding to $\Delta=500$ in the top two rows, split into cluster-scale (top row) and galaxy-scale (second row, from top) masses, and $\Delta=200$ in the bottom two rows, split into cluster-scale (third row, from top) and galaxy-scale (bottom row) masses. Within each box, each colour varies the velocity anisotropy, $\beta$.}
    \label{Fig - aperture velocity dispersion profiles - varying inner slope}
\end{figure*}

\section{3. \quad Analysis}
\vspace{2mm}
\subsection{3.1. \quad Analytical profiles for the ideal physical halos}

To constrain $\xi$ in these four regimes, the aperture velocity dispersion profiles must be analytically modelled for the ideal physical halos in terms of the parameters outlined. For this analysis, we model the ideal physical halos with the scale-free density and gravitational potential profiles detailed in Equations \eqref{ideal physical halo profile} and \eqref{ideal physical halo gravitational potential}, respectively. With these profiles, we can analytically predict and numerically trace the radial, line of sight and subsequently the aperture velocity dispersion profiles for these halos in scale-free form by making use of the kinematic profiles detailed in Section 2.3. 

\subsubsection{The radial velocity dispersion of the ideal physical halos}

The radial velocity dispersion profiles for the ideal physical halos, in scale-free form, are derived from the general solution of the Jeans equation from Equation \eqref{scale-free radial velocity dispersion}, taking on the analytic form:
\begin{equation}\label{ideal physical halo radial velocity dispersion}
\begin{aligned}
    \frac{\sigma _\mathrm{r}^2(s, c, \alpha, \beta )}{v_\mathrm{vir}^2}&= u(c, \alpha) \cdot s^{\alpha - 2\beta }(1+cs)^{3 - \alpha } \\
    &\times \int _s^\infty \frac{s^\prime {}^{2\beta - \alpha - 2}\mathrm{d}s^\prime }{(1+cs^\prime )^{3 - \alpha }}\left[\int _0^{s^\prime }  \frac{s^\prime {}^\prime {}^{2 - \alpha }\mathrm{d}s^\prime {}^\prime }{(1+cs^\prime {}^\prime )^{3 - \alpha}}\right].
\end{aligned}
\end{equation}
These scale-free radial velocity dispersion profiles, $\sigma_\mathrm{r}/v_\mathrm{vir}$, are shown in Figure \ref{Fig - radial velocity dispersion profiles}, as a function of the dimensionless halocentric radius, $s\equiv r/r_\mathrm{vir}$. 
These panels encompass the variation of the halo's kinematics within the desired parameter space of $\alpha \in [0, 1.5]$ and $\beta \in [0, 0.5]$, and at fixed concentrations, $c$, corresponding to the halo mass scales in the desired regimes. Within these panels, as the velocity anisotropy is increased from $\beta = 0$ to $\beta = 0.5$ from left to right across each row, the more cored, $\alpha =0$ and $\alpha=0.5$, halo profiles become increasingly divergent at small halo radii. However, in all instances, the isotropic, $\beta = 0$, profiles remain well-behaved and bounded for all halo inner slopes. 

\subsubsection{The surface mass density of the ideal physical halos}

To predict the projected velocity dispersion profiles for the ideal physical halos, the surface mass density profile must be modelled. By Equation \eqref{scale-free surface density}, we can express this projected density profile, in scale-free form, as:
\begin{equation}\label{ideal physical halo surface mass density}
    \frac{\Sigma (S, c, \alpha)}{\Sigma _\mathrm{vir}}=\frac{u(c, \alpha)}{2} \cdot \int _{S}^\infty \frac{s^{1-\alpha  }\mathrm{d}s}{(1+cs)^{3 - \alpha  }\sqrt{s^2-S^2}}.
\end{equation}

\subsubsection{The line of sight velocity dispersion of the ideal physical halos}

From the derived radial velocity dispersion and surface mass density profiles for the ideal physical halos, the corresponding line of sight velocity dispersion can be derived from Equation \eqref{scale-free los velocity dispersion}, producing the scale-free expression:
\begin{equation}\label{ideal physical halo los velocity dispersion}
\begin{aligned}
    \frac{\sigma _\mathrm{los} ^2(S, c, \alpha, \beta)}{v_\mathrm{vir}^2}&=u(c, \alpha) \cdot  \int _{S}^\infty \Biggl\{ \frac{(1- \frac{\beta S^2}{s^2})s^{1-2\beta }\mathrm{d}s}{\sqrt{s^2-S^2}} \\
    &\times \frac{\int _s^\infty \frac{s^\prime {}^{2\beta - \alpha -2}\mathrm{d}s^\prime }{(1+cs^\prime )^{3 - \alpha }} \left[\int _0^{s^\prime} \frac{s^\prime {}^\prime {}^{2-\alpha  }\mathrm{d}s^\prime{}^\prime }{(1+cs^\prime {}^\prime )^{3 - \alpha  }}\right] }{\int _{S}^\infty \frac{s^{1-\alpha  }\mathrm{d}s}{(1+cs)^{3 - \alpha  }\sqrt{s^2-S^2}}} \Biggr\}.
\end{aligned}
\end{equation}
These scale-free line of sight velocity dispersion profiles, $\sigma_\mathrm{los}/v_\mathrm{vir}$, are shown in Figure \ref{Fig - los velocity dispersion profiles}, as a function of the dimensionless projected radius, $S\equiv R/r_\mathrm{vir}$. 

\subsubsection{The aperture velocity dispersion of the ideal physical halos}

Taking the profiles for the surface mass density and the line of sight velocity dispersion into Equation \eqref{scale-free aperture velocity dispersion}, the aperture velocity dispersion profile for the ideal physical halos, in scale-free form, takes the form:
\begin{equation}\label{ideal physical halo aperture velocity dispersion}
\begin{aligned}
    \frac{\sigma _\mathrm{ap} ^2(<S_\mathrm{ap}, c, \alpha, \beta)}{v_\mathrm{vir}^2}&=u(c, \alpha)\cdot \frac{\int _0 ^{S_\mathrm{ap}} S \mathrm{d}S  \Biggl\{ \int _{S }^\infty  \frac{(1- \frac{\beta S {}^2}{s^2})s^{1-2\beta }\mathrm{d}s}{\sqrt{s^2-S {}^2}} }{\int _0 ^{S_\mathrm{ap}} S \mathrm{d}S  \left [\int _{S }^\infty \frac{s^{1-\alpha  }\mathrm{d}s}{(1+cs)^{3 - \alpha  }\sqrt{s^2-{S {}^2}}}\right]}  \\ 
     & \hspace{-10mm}\times   \int _s^\infty \frac{s^\prime {}^{2\beta - \alpha -2}\mathrm{d}s^\prime }{(1+cs^\prime )^{3 - \alpha }} \left[\int _0^{s^\prime} \frac{s^\prime {}^\prime {}^{2-\alpha  }\mathrm{d}s^\prime{}^\prime }{(1+cs^\prime {}^\prime )^{3 - \alpha  }}  \right]  \Biggr\}.
\end{aligned}
\end{equation}
These scale-free aperture velocity dispersion profiles, $\sigma_\mathrm{ap}/v_\mathrm{vir}$, are shown in Figure \ref{Fig - aperture velocity dispersion profiles}, as a function of the dimensionless aperture radius, $S_\mathrm{ap} \equiv R_\mathrm{ap}/r_\mathrm{vir}$. 

\subsection{3.2. \quad Constraints on $\xi$}

When the scale-free aperture velocity dispersion, $\sigma_\mathrm{ap}/v_\mathrm{vir}$, of the ideal physical halos from Equation \eqref{ideal physical halo aperture velocity dispersion} is traced within a given dimensionless aperture radius, $S_\mathrm{ap} \equiv R_\mathrm{ap}/r_\mathrm{vir}$, this will evaluate the dimensionless parameter $\xi \equiv \sigma_\mathrm{ap} (<R_\mathrm{ap})/ v_\mathrm{vir}$. By measuring this value globally over the parameter space corresponding to those outlined within each of the four regimes, the values of $\xi_\mathrm{min}$ and $\xi_\mathrm{max}$ in each case will be determined. 

\subsubsection{Constraints on $\xi$ within each of the four regimes}

These four regimes in which $\xi$ is to be bounded can be traced within the panels of Figure \ref{Fig - aperture velocity dispersion profiles}: as the region corresponding to $R_\mathrm{ap}/r_\mathrm{vir} \in [0.2, 2]$, when $c=2.5$, in the top row, and $c=5$, in the middle row; and for $R_\mathrm{ap}/r_\mathrm{vir}  \in [0.1, 1]$, when $c=5$, in the middle row, and $c=10$, in the bottom row. It is clear from this figure that the values of $\xi_\mathrm{min}$ and $\xi_\mathrm{max}$ in each of these four regimes are set at the end-points of the range in aperture radii, in each instance. Figure \ref{Fig - aperture velocity dispersion profiles - varying inner slope} evaluates these $\xi \equiv \sigma_\mathrm{ap} (<R_\mathrm{ap}) /v_\mathrm{vir}$ profiles from Figure \ref{Fig - aperture velocity dispersion profiles}, now at each of these end-points in aperture radii, within each of the four regimes, producing eight distinct windows. 
The values of $\xi_\mathrm{min}$ and $\xi_\mathrm{max}$ are then simply the minimum and maximum values within each of the four rows of Figure \ref{Fig - aperture velocity dispersion profiles - varying inner slope}. These results produce the constraints in Table \ref{Table - xi bounds}.

\section{4. \quad Results}
\vspace{2mm}
\subsection{4.1. \quad The $M_\mathrm{vir} - \sigma_\mathrm{ap}$ scaling relations}

With the bounds devised for $\xi_\mathrm{min}$ and $\xi_\mathrm{max}$, as detailed in Table \ref{Table - xi bounds}, the halo mass scaling relations from Section 2.4 --- Equations \eqref{M200 mass bound} and \eqref{M500 mass bound} --- will be constrained in each of the four outlined regimes. \textcolor{black}{These constraints allow us to form a quantified prediction for each scaling relation, along with a quantified uncertainty, with this uncertainty taken to encompass the total range predicted for each scaling relation, propagated by these constraints in $\xi$, within each regime. }

\begin{figure*}[h!]
    \centering
    \includegraphics[width=\textwidth]{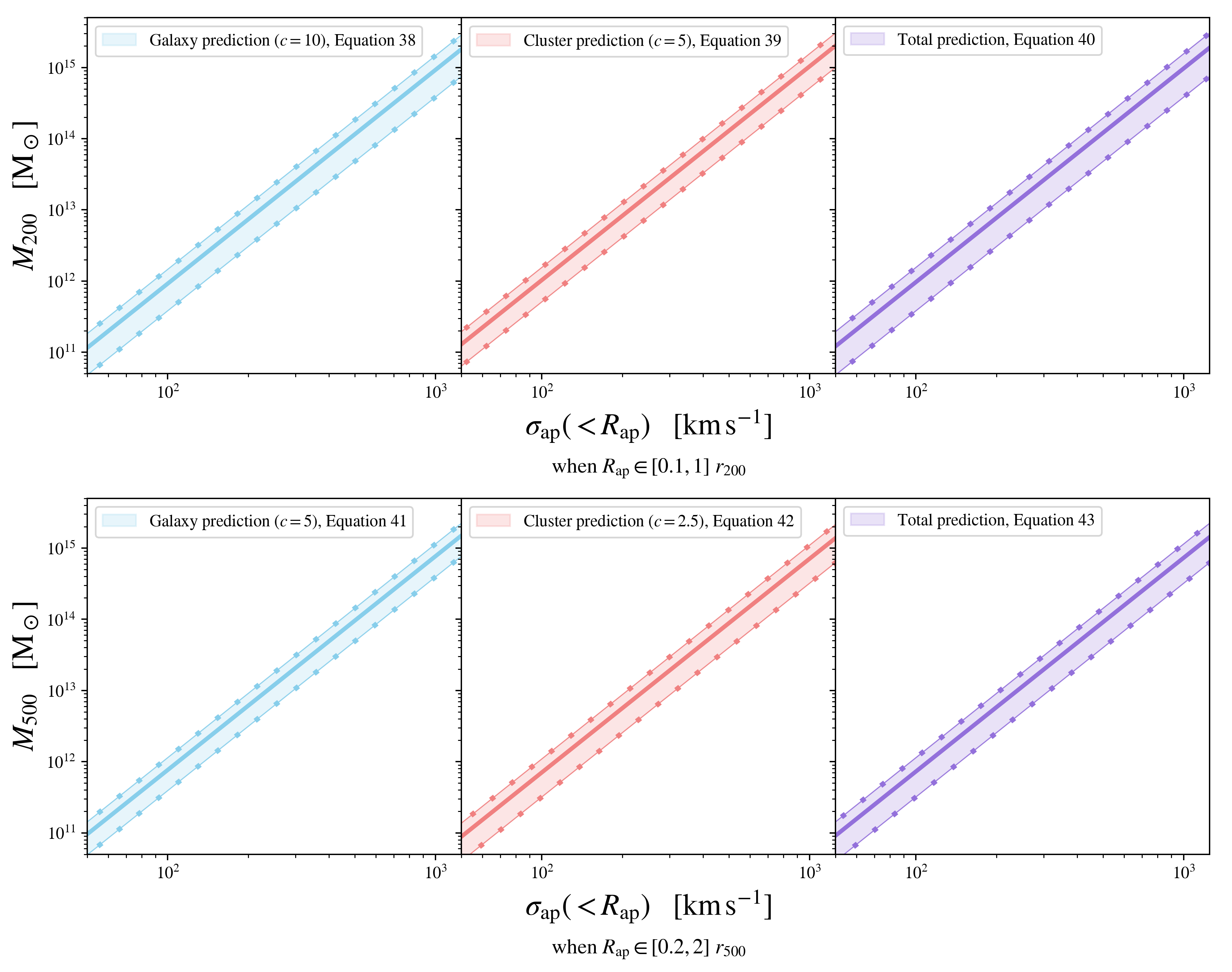}
    \caption{Our predictions for the halo mass - aperture velocity dispersion scaling relations, $M_{200} - \sigma_\mathrm{ap}$ and $M_{500} - \sigma_\mathrm{ap}$, when the aperture velocity dispersion is measured within an aperture radius, $R_\mathrm{ap}$, inside the range specified in each panel. The uncertainties in these scaling relations are quantified by constraints in the dimensionless parameter $\xi$, given in Table \ref{Table - xi bounds}, derived over specified halo parameters. These scaling relations are presented for galaxy halo masses (left, in light blue) and cluster halo masses (middle, in light red), and then combined (right, in purple) to make a total prediction, encompassing both scales. The solid dotted lines enclosing each interval correspond to the minimum and maximum bounds in the scaling relation, with the solid central line tracing its mid-range value. }
    \label{Fig - virial mass scaling relations}
\end{figure*}

These four scaling relations are shown in Figure \ref{Fig - virial mass scaling relations}, with the $M_{200} - \sigma_\mathrm{ap}$ scaling relations shown in the top row, and the $M_{500} - \sigma_\mathrm{ap}$ scaling relations shown in the bottom row, each at galaxy and cluster masses, as in the left-most and central panels, respectively.  
In the right-most panels of this figure, as shown in purple, the galaxy and cluster mass predictions are merged to form a total, encompassing prediction for each scaling relation. Each of these scaling relations and their predicted uncertainties are presented analytically below.

\subsubsection{The $M_{200} - \sigma_\mathrm{ap}$ scaling relation}

Taking the mid-range value of the $M_{200} - \sigma_\mathrm{ap}$ scaling relation and the uncertainty encompassing its bounding, this scaling relation is predicted, at galaxy-scale halo masses, as:
\begin{equation}\label{M200 scaling relation - galaxy-scale}
\begin{aligned}
    M_{200} &= \left(9.21 \pm 5.39\right) \cdot  \left[\frac{\sigma_\mathrm{ap}}{100\,\mathrm{km\, s^{-1}}}\right]^3 \cdot 10^{11} \, \mathrm{M}_\odot,
\end{aligned}
\end{equation}
and at cluster-scale halo masses, as:
\begin{equation}\label{M200 scaling relation - cluster-scale}
\begin{aligned}
    M_{200} &= \left(10.32 \pm 5.21\right) \cdot  \left[\frac{\sigma_\mathrm{ap}}{1000\,\mathrm{km\, s^{-1}}}\right]^3 \cdot 10^{14} \, \mathrm{M}_\odot,
\end{aligned}
\end{equation}
each with a corresponding uncertainty of $58.5\%$ and $50.5\%$, respectively. Merging these intervals, as in Figure \ref{Fig - virial mass scaling relations}, the total prediction is then:
\begin{equation}\label{M200 scaling relation - combined}
\begin{aligned}
    M_{200} &= \left(9.68 \pm 5.86\right) \cdot  \left[\frac{\sigma_\mathrm{ap}}{1000\,\mathrm{km\, s^{-1}}}\right]^3 \cdot 10^{14} \, \mathrm{M}_\odot,
\end{aligned}
\end{equation}
with the uncertainty increasing to $60.5\%$. 

\subsubsection{The $M_{500} - \sigma_\mathrm{ap}$ scaling relation}

Similarly, when taking the mid-range value and the associated uncertainty for the $M_{500} - \sigma_\mathrm{ap}$ scaling relation, we deduce that at galaxy-scale halo masses, this scaling relation is predicted as:
\begin{equation}\label{M500 scaling relation - galaxy-scale}
\begin{aligned}
    M_{500} &= \left(7.65 \pm 3.73\right) \cdot  \left[\frac{\sigma_\mathrm{ap}}{100\,\mathrm{km\, s^{-1}}}\right]^3 \cdot 10^{11} \, \mathrm{M}_\odot,
\end{aligned}
\end{equation}
and at cluster-scale halo masses, as:
\begin{equation}\label{M500 scaling relation - cluster-scale}
\begin{aligned}
    M_{500} &= \left(7.06 \pm 3.87\right) \cdot  \left[\frac{\sigma_\mathrm{ap}}{1000\,\mathrm{km\, s^{-1}}}\right]^3 \cdot 10^{14} \, \mathrm{M}_\odot,
\end{aligned}
\end{equation}
each with a corresponding uncertainty of $48.7\%$ and $54.8\%$, respectively. Upon merging these intervals, the total prediction is then:
\begin{equation}\label{M500 scaling relation - combined}
\begin{aligned}
    M_{500} &= \left(7.29 \pm 4.09\right) \cdot  \left[\frac{\sigma_\mathrm{ap}}{1000\,\mathrm{km\, s^{-1}}}\right]^3 \cdot 10^{14} \, \mathrm{M}_\odot,
\end{aligned}
\end{equation}
with the uncertainty increasing to $56.2\%$.

\subsection{4.2. \quad Parameter dependence of the scaling relations}

To derive these scaling relations, four parameters have been modelled: the aperture radius, $R_\mathrm{ap}$, the halo concentration, $c$, the halo density inner slope, $\alpha$, and the velocity anisotropy, $\beta$. With reference to Figure \ref{Fig - aperture velocity dispersion profiles - varying inner slope}, tracing $\xi \equiv \sigma_\mathrm{ap}(<R_\mathrm{ap})/v_\mathrm{vir}$ throughout this parameter space, we can comment on the dependence of the values of $\xi_\mathrm{min}$ and $\xi_\mathrm{max}$ --- and subsequently the dependence of the scaling relations --- on these assumed parameters.

\subsubsection{Dependence on the aperture radius $R_\mathrm{ap}$}

When comparing our predictions to real observations, the aperture radius, $R_\mathrm{ap}$, is the most variable and observationally-sensitive parameter assumed in this model: as its value chosen in a given observation will depend upon the group statistics available for each galaxy or galaxy cluster. In our model, we circumvented this sensitivity by prescribing a range in values for the dimensionless aperture, in scale-free form $S_\mathrm{ap} \equiv R_\mathrm{ap}/r_\mathrm{vir}$, to reasonably capture this variability. 

Expectedly, within Figure \ref{Fig - aperture velocity dispersion profiles - varying inner slope} it is clear that at smaller aperture radii (as in the left-most column) there is a larger difference between each constant-$\beta$ curve tracing $\xi$ at any halo inner slope, $\alpha$, than there is at larger aperture radii (as in the right-most column). As these aperture velocity dispersion orbits become more strongly convergent within larger aperture sizes, it is clear that choosing a higher minimum aperture radius in this model would pose a stronger constraint on the halo mass in each scaling relation. Even more, if the aperture radius could be chosen at a large, fixed value of $R_\mathrm{ap} \simeq r_{200} \simeq 2 r_{500}$, the uncertainty in the scaling relation would be significantly reduced; however, this would reduce the applicability and generality of our predictions. Furthermore, observations usually do not know the size of the aperture radius in relation to the halo's virial radius, in either $r_{200}$ or $r_{500}$ units, further complicating assuming a fixed aperture radius in any realistic prediction. 

\subsubsection{Dependence on the concentration parameter $c$}

Throughout this study, we have fixed the halo concentration, $c$, to particular halo mass scales: $c=5$ and $c=10$ for cluster-scale and galaxy-scale halo masses, respectively, when $\Delta = 200$; and $c=2.5$ and $c=5$ for cluster-scale and galaxy-scale halo masses, respectively, when $\Delta = 500$. In Figure \ref{Fig - virial mass scaling relations}, showing the predicted scaling relations at different halo mass scales --- and hence different concentrations --- it can be seen that there is only a weak sensitivity in these relations to this choice in concentration. Comparing the galaxy-scale and cluster-scale scaling relations, in either $\Delta=200$ or $\Delta=500$ convention, it is clear that the intervals spanned in these predictions are strongly overlapping. Taking the $M_{200} - \sigma_\mathrm{ap}$ scaling relation, where a change from $c=5$ to $c=10$ represents a change in halo mass of approximately three orders of magnitude, the fact that these separate regimes are strongly overlapping implies that these predictions are not strongly sensitive to the choice of concentration, or correspondingly to the halo mass scale. 

In $\Lambda \mathrm{CDM}$ cosmological simulations, concentrations of dark matter halos typically fall within $c \in [4, 22]$ \autocite[e.g.][]{Zhao2003}. Despite not considering, when $\Delta=200$, the concentrations of super-massive clusters, $c\lesssim 5$, or dwarf galaxies, $c\gtrsim 10$, as this sensitivity to the concentration is weak, it is expected that these limiting halo mass scales will be reasonably contained within our predictions.

\subsubsection{Dependence on the halo inner slope $\alpha$}

The dependence of the value of $\xi$ on the halo inner slope, $\alpha$, is evident in Figure \ref{Fig - aperture velocity dispersion profiles - varying inner slope}, tracing $\xi$ continuously between $\alpha=0$ and $\alpha=1.5$ in each of the four regimes that bound the halo mass scaling relation. Within these panels, in the bottom three rows, the rate of change of $\xi$ with respect to $\alpha$ (the gradient of the curves in these panels) is consistently maximised at the largest, cuspiest halo inner slope of $\alpha=1.5$ in this range. Consequently, in these regimes, the value of $\xi_\mathrm{max}$ is strongly sensitive to the maximum value chosen for $\alpha$: if a larger maximum halo inner slope $\alpha \gtrsim 1.5$ was to be numerically or observationally motivated, the lower bound of these scaling relations would decrease, increasing the total uncertainty in the estimate. Alternatively, and physically more likely, if cuspy halo inner slopes could be ruled out below some value $\alpha \lesssim 1.5$, this would decrease the uncertainty in these predictions. 

In contrast, in the top $c=2.5$ row of Figure \ref{Fig - aperture velocity dispersion profiles - varying inner slope}, this trend changes: the rate of change of $\xi$ with respect to $\alpha$ becomes steepest toward the minimum, cored halo inner slope, $\alpha=0$, particularly for the most anisotropic, $\beta=0.5$, curve. In this case, $\xi_\mathrm{min}$ is strongly dependent on this choice in the minimum value of $\alpha$. However, as the existence of cores remains unresolved \autocite[see, e.g.][for potential interpretations]{Oman2015}, and as halo inner slopes $\alpha < 0$ are unphysical, this sensitivity to the minimum value of $\alpha$ in our model is unlikely to be exploited, without strong precedent to rule out the possibility of halo cores. As such, it is the maximum bound of the halo's inner slope, chosen in our analysis as $\alpha = 1.5$, that imposes the most significant dependence for our predictions, across all four chosen regimes.

\subsubsection{Dependence on the velocity anisotropy $\beta$}

As seen in Figure \ref{Fig - aperture velocity dispersion profiles - varying inner slope}, across all four intervals, increasing the value of the velocity anisotropy, $\beta$, always increases the value of $\xi$ at any halo inner slope between $\alpha =0$ and $\alpha = 1.5$. In particular, when all other parameters are fixed, it is always the isotropic, $\beta = 0$, value that sets the minimum value of $\xi$, and the most anisotropic, $\beta = 0.5$, value that sets the maximum value of $\xi$. As such, the values of $\xi_\mathrm{min}$ and $\xi_\mathrm{max}$ in the scaling relations are sensitive to both the minimum and maximum bounds for $\beta$, respectively.

More detailed analysis of the kinematics of a dark matter halo could choose to model the anisotropy by some function $\beta = \beta (r)$ of halocentric radius $r$. One well-known example is the Osipkov-Merritt (OM) velocity anisotropy model \autocite{Merritt1985}, which models the anisotropy as isotropic below the so-called anisotropy radius, and radially increasing beyond. Analytic studies have shown that the scale-free aperture velocity dispersion, $\sigma_\mathrm{ap}/v_\mathrm{vir}$, for an NFW profile modelled with an OM anisotropy, is strongly converged to the same profile with an isotropic, $\beta=0$, spherically-averaged anisotropy \autocite{LokasMamon2001}. As such, if the OM anisotropy model or a similar radially increasing model could be observationally motivated --- rather than assuming a parameter space of spherically-averaged, constant $\beta$ values, as in this study --- such modelling would be expected to significant decrease the uncertainty within the predicted scaling relations.

\subsection{4.3. \quad Limitations of the model}

\textcolor{black}{
The results derived in this study are limited in their application to observable systems by the number of accessible kinematic tracers of the gravitational potential (e.g. satellite galaxies). The number of satellites increases from of order unity ($\mathcal{O}(1)$) on Milky Way mass scales to $\mathcal{O}(10^2-10^3)$ on the scale of galaxy groups and clusters \autocite[see, e.g.][]{Berlind.Weinberg.2002}{}{}, and so it is on the scales of massive galaxy groups and clusters that radial varying velocity dispersions derived directly from a system's satellite galaxies can be measured most accurately. When limited by poor statistics, the observed aperture velocity dispersion will not necessarily converge with its analytic prediction, even if the halo is otherwise self-similar. This incomplete sampling of a halo's kinematic tracers imposes the need for an additional uncertainty, which would need to be introduced into the scaling relations to permit a more reasonable application to observations \autocite[see, e.g.][]{Robotham2011}{}{}. Whilst not pursued here, the contribution of this statistical error could be investigated with numerical simulations, to better inform this applicability of our results.}

\textcolor{black}{
Beyond the practical challenge of having a sufficient number of kinematic tracers, 
the hierarchical nature of structure formation means that the assumptions of virial equilibrium and spherical symmetry of the gravitational potential are not always sufficient. Numerical simulations reveal that halos tend to be preferentially prolate, with more-massive haloes tending to be least spherical and most prolate \autocite[e.g.][]{Bett.2007}{}{}; therefore we expect the measured line of sight velocity distribution to be biased, resulting in projection effects that will cause deviations from our predicted scaling relations, otherwise not accounted for in our model. Similarly, the prevalence of accretion and mergers impacts the assumption of virial equilibrium \autocite[e.g.][]{Power.2012}{}{}, causing deviations from the expected equilibrium velocity dispersion for of order a dynamical time. Numerical simulations are required for a careful treatment of these effects. }

\section{5. \quad Conclusion}

This study has explored the relationship between the dark matter halo mass, its density profile and the kinematic properties of its tracers moving within its gravitational potential. In particular, we have demonstrated that an analysis of the kinematic profiles of an idealised framework of halos --- referred to as the ideal physical halos --- can be used to make predictions for the scaling relation between the halo's mass and its observed aperture velocity dispersion, with this method entirely independent of numerical models or simulation calibrations. 

Through this approach, we have predicted the $M_{200} - \sigma_\mathrm{ap}$ scaling relation with an uncertainty of $60.5\%$, and the $M_{500} - \sigma_\mathrm{ap}$ scaling relation with an uncertainty of $56.2\%$, with both of these predictions accounting for variation in the aperture radius, the halo mass scale, the halo's inner slope and the velocity anisotropy of its tracer populations. The uncertainty in these results are approximately an order of magnitude below the estimators utilised in current spectroscopic surveys targeting the halo mass, as calibrated from numerical group catalogues. The implication of our study is that the variation of the halo's structural and kinematic profiles contribute only a small component of the uncertainty established in these numerical calibrations. Larger uncertainty contributions to this scaling relation are expected to arise in projection effects when spherical symmetry is broken, and in sampling biases when only few group members are observed in a given halo. This latter effect introduces a significant statistical error when compared to our theoretical predictions, as the aperture velocity dispersion measured when only few tracers are available can attain a large deviation from its analytically predicted value. If future studies robustly quantify a sampling error to couple to our analytic predictions, this would produce a more accessible halo mass estimate for application to present and upcoming spectroscopic surveys. 

More fundamentally, our study has developed an analytic tool-kit for studying the properties of dark matter halos in terms of well-defined parameters: grounded upon physically valid assumptions and circumventing any universality in halo form, albeit assuming spherical symmetry, and by prescribing physically-motivated ranges for each parameter value.  Within this parameter space, we showed that our scaling relation predictions are most sensitive to the minimum aperture radius in which the aperture velocity dispersion is measured, the maximum value permitted for the halo's inner slope, taken in our study as $\alpha=1.5$, and the minimum and maximum bounds in the velocity anisotropy, taken in our study as $\beta = 0$ and $\beta = 0.5$. The dependence of these scaling relations on the halo's concentration was shown to be minimal, implying a weak dependence of this estimator on the halo's mass scale. 

As complex and dynamic objects, dark matter halos have often remained in the exclusive focus of simulations. We have demonstrated that an analytical framework can be constructed for an idealised class of halos, to predict the halo mass - aperture velocity dispersion scaling relation in a way that is theoretically grounded, simulation-independent and largely insensitive and agnostic to any universality in the halo's structural and kinematic profiles. We hope that our predictions will assist toward a strong and reliable halo mass estimator for upcoming probes of the Halo Mass Function, which excites the prospect of constraining the nature of dark matter.

This paper is the first in a series of papers predicting the scaling relations of dark matter halos. The second paper will predict the scaling relations of galaxy clusters with observable measures of its intracluster gas emission: in particular, its mean-weighted X-ray temperatures and the integrated Sunyaev-Zeldovich effect. \textcolor{black}{Future work will use cosmological hydrodynamical simulations \autocite[e.g., The Three Hundred][]{Cui.2018}{}{} to test the range of validity of this model's predictions on astrophysically realistic systems.}

\vspace{1cm}

 AS acknowledges the support of the Australian Government Research Training Program Fees Offset; the Bruce and Betty Green Postgraduate Research Scholarship; and The University Club of Western Australia Research Travel Scholarship. AS and CP acknowledge the support of the ARC Centre of Excellence for All Sky Astrophysics in 3 Dimensions (ASTRO 3D), through project number CE170100013.

 CB gratefully acknowledges support from the Forrest Research Foundation.

\printbibliography

\end{document}